\documentclass[aps,prd,onecolumn,groupedaddress,showpacs,nofootinbib,amssymb]{revtex4}

\usepackage[dvips]{graphicx}
\usepackage{amssymb}
\usepackage{amsmath}
\usepackage{graphicx,color}
\usepackage{amsfonts}
\usepackage{bm}
\usepackage{cancel}
\usepackage{comment}
\newcommand{\be}{\begin{equation}}
\newcommand{\ee}{\end{equation}}
\newcommand{\bea}{\begin{eqnarray}}
\newcommand{\eea}{\end{eqnarray}}
\newcommand{\beaa}{\begin{eqnarray*}}
\newcommand{\eeaa}{\end{eqnarray*}}

\newcommand{\nn}{\nonumber \\}
\newcommand{\e}{\mathrm{e}}

\allowdisplaybreaks[4]
\begin{document}

\title{Ghost-free non-local $F(R)$ Gravity Cosmology}
\author{Shin'ichi~Nojiri,$^{1,2}$\,\thanks{nojiri@gravity.phys.nagoya-u.ac.jp}
S.~D.~Odintsov,$^{3,4}$\,\thanks{odintsov@ieec.uab.es}
V.~K.~Oikonomou,$^{5,6}$\,\thanks{v.k.oikonomou1979@gmail.com} }
\affiliation{ $^{1)}$ Department of Physics, Nagoya University,
Nagoya 464-8602, Japan \\
$^{2)}$ Kobayashi-Maskawa Institute for the Origin of Particles
and the Universe, Nagoya University, Nagoya 464-8602, Japan \\
$^{3)}$ ICREA, Passeig Luis Companys, 23, 08010 Barcelona, Spain\\
$^{4)}$ Institute of Space Sciences (IEEC-CSIC) C. Can Magrans
s/n, 08193 Barcelona, Spain\\
$^{5)}$ Department of Physics, Aristotle University of
Thessaloniki, Thessaloniki 54124, Greece\\
$^{6)}$ International Laboratory for Theoretical Cosmology, Tomsk
State University of Control Systems and Radioelectronics (TUSUR),
634050 Tomsk, Russia}

\begin{abstract}
In this work we shall study ghost-free non-local $F(R)$ gravity
models. Firstly we shall demonstrate how the ghost degrees of
freedom may occur in the non-local $F(R)$ gravity models, and
accordingly we shall modify appropriately the gravitational action
of non-local $F(R)$ gravity models in order to eliminate the
ghosts. Also we shall investigate how the (anti-)de Sitter and the
Minkowski spacetime cosmological solutions may arise in the
theory, and we investigate when these solutions are stable.
Moreover, we shall examine the inflationary phenomenology of the
Jordan frame ghost-free non-local $F(R)$ gravity. We shall study
two $F(R)$ gravity models, the power law $F(R)$ gravity model
$\sim R^n$ with $1<n<2$, $n\neq 2$ and the $R^2$ model, assuming
that the slow-roll condition holds true for the Hubble rate during
the inflationary era $\dot{H}\ll H^2$ and that the general
constant-roll condition $\ddot{\phi}=3\beta H\dot{\phi}$ holds
true for the evolution of the scalar field, which includes the
slow-roll case for $\beta=0$. As we shall demonstrate, the
power-law non-local $F(R)$ gravity case can produce a viable
inflationary era, compatible with observations, in the
constant-roll case, unlike for the $R^2$ model. It is conceivable
that the results are model dependent, as in the ordinary vacuum
$F(R)$ gravity.
\end{abstract}

\maketitle

\section{Introduction}

Einstein's final task and idea was to find a unified field theory
of everything, a theory in which all the fundamental interactions
might be explained in the same framework in a fundamental way.
Ever since, theoretical physicists focused in finding the
fundamental theory of everything, with the most important theory
that emerged being, to our opinion, string theory quantified in
its various $M$-theory variants. However, the $M$-theory
predictions provide phenomenological results that can be verified
at very high energies, making the experimental testing at a
pragmatic level, rather an far future or a futile task. Thus the
quantization of gravity remains an unsolved problem, and by
quantization we mean the whole process of embedding the
gravitational field in the already successful field theories that
the quantization procedure resulted to experimental verifications
of their predictions.

In the process of finding the fundamental theory, however, many
physicists realized that it may be possible to discover imprints
of the unified theory of everything in the classical physics, by
using a up-bottom approach, contrary to the string theory
framework which is a bottom-up theory, trying to extract the
classical theory, from the fundamental theory itself. Indeed, if
one adopts the up-bottom approach, one may seek for effective
gravitational theories, that may include terms of the quantum
unified theory of everything, accompanying the standard classical
theory of the Einstein-Hilbert gravity. The reason for trying to do
this is mainly because, although the classical theory of gravity
seems to be perfectly describing astrophysical phenomena, at large
scales it seems to be lacking of a self-consistent description.
Indeed, after the discovery of the accelerating expansion of the
Universe
\cite{Perlmutter:1998np,Riess:1998cb,Spergel:2003cb,Spergel:2006hy,
Komatsu:2008hk,Komatsu:2010fb,Tegmark:2003ud,Seljak:2004xh,Eisenstein:2005su,
Jain:2003tba}, it was realized that a modification of gravity is
needed in order to perfectly describe the late-time acceleration
era. In this line of research, many modified gravity models have
been proposed
\cite{Nojiri:2017ncd,Nojiri:2010wj,Nojiri:2006ri,Capozziello:2011et,
Capozziello:2010zz,delaCruzDombriz:2012xy,Olmo:2011uz},
in the context of which the late-time acceleration may be
described, or sometimes inflation and dark energy may be described
in a unified way within the same theoretical framework
\cite{Nojiri:2003ft}.

One of the most promising modified gravity theories, which also
has a direct with an underlying quantum effective theory of
gravity, is the non-local gravity, see
Refs.~\cite{Modesto:2017sdr,Belgacem:2017cqo,Koshelev:2016xqb} for
recent reviews on the subject. This theory was preliminary
proposed in Ref.~\cite{Wetterich:1997bz}, however the original
version could not take into account cosmological phenomena. A
consistent non-local gravity model that took also into account
cosmological phenomena, was proposed in \cite{Deser:2007jk}, and
ever since it has been actively investigated
\cite{Nojiri:2007uq,ArkaniHamed:2002fu,Nojiri:2010pw,Joukovskaya:2007nq,Calcagni:2007ef,
Jhingan:2008ym,Capozziello:2008gu,Koshelev:2008ie,Nesseris:2009jf,
Deffayet:2009ca,Calcagni:2009dg,Cognola:2009jx,Bronnikov:2009az,Calcagni:2010ab,
Vernov:2010ui,Barnaby:2010kx,Dimitrijevic:2019pct,Dialektopoulos:2018iph,Calmet:2018rkj,Bahamonde:2017sdo,Elizalde:2011su,Bamba:2012ky,Deser:2013uya,Deser:2019lmm,Maggiore:2014sia,
Chan:2012jj,Koshelev:2016xqb,Zhang:2011uv}. In general, the
non-local models of gravity, and also of non-local $F(R)$ gravity,
can be rewritten in a local form by introducing scalar field(s).
With the present paper we aim to demonstrate the non-local $F(R)$
theories of gravity are compromised by the existence of ghost
degrees of freedom, that may result to destabilizing the effective
gravitational theory framework of non-local $F(R)$ gravity. In
addition and more importantly, we aim to propose a direct remedy
for the non-local $F(R)$ gravity, by providing a ghost-free
modification of the non-local $F(R)$ gravity theory. We shall
thoroughly investigate the cosmological aspects of these theories,
and we shall demonstrate that it is possible for these theories to
have an exact solution which describe the (anti-)de Sitter or the
flat Minkowski spacetime. We also find the conditions for which
the solutions become stable. Finally, we shall thoroughly
investigate the inflationary phenomenological aspects of the
ghost-free non-local $F(R)$ gravity theory in the Jordan frame,
for a flat cosmological geometry, by assuming a slow or a
constant-roll evolution for the scalar field $\ddot{\phi}=3\beta
H\dot{\phi}$, and the standard slow-roll approximation for the
Hubble rate, namely $\dot{H}\ll H^2$, during the whole
inflationary era. We examine two cases of power-law $F(R)$ gravity
models, a standard power-law $\sim R^n$, with $1<n<2$ and $n\neq
2$, and a $R^2$ gravity model. The reason of discriminating these
two cases is mainly a technical simplification that occurs in the
$R^2$ case, which enables us to extract more accurate results
analytically. For both the cases we present the technical
formalism for calculating the slow-roll indexes and the
corresponding spectral index of the primordial curvature
perturbations and the tensor-to-scalar ratio, and we calculate in
detail the Hubble rate as a function of the $e$-foldings number
and the free parameters of the theory. As we demonstrate, the
power-law $F(R)$ gravity can be compatible with the latest (2018)
Planck constraints on the inflationary parameters
\cite{Akrami:2018odb}, if the scalar field  satisfies the
constant-roll evolution, even for small $\beta$ values, however,
the $R^2$ produces less appealing phenomenological results.

This paper is organized as follows: In section II we demonstrate
how ghosts may occur in non-local $F(R)$ gravity theories, while
in section III we discuss how we can modify the non-local $F(R)$
gravity models action in order to avoid the ghost degrees of
freedom. In section IV we study how the (anti-)de Sitter and the Minkowski
spacetime solutions can be realized by the ghost free non-local
$F(R)$ gravity model, and in section V we investigate the
phenomenological inflationary predictions of the non-local $F(R)$
gravity models, by working in the Jordan frame and in the absence
of any matter perfect fluids. We study the ghost-free power-law
and $R^2$ non-local $F(R)$ and we confront the results with the
latest Planck data. Finally, the conclusions follow in the end of
the paper.

\section{Ghost in Non-local $F(R)$ Gravity}

Before constructing the ghost-free models of non-local gravity, we
shall discuss how the ghosts occur in the standard non-local
$F(R)$ models. We may consider the non-local $F(R)$ gravity, whose
action is given by,
\begin{equation}
\label{nlFR1}
S= \int d^4 x \sqrt{-g} \left\{ \frac{1}{2\kappa^2} \left( R + F(R) \Box^{-k} R\right)
+ \mathcal{L}_\mathrm{matter} \left( g_{\mu\nu}, \Phi_i \right) \right\} \, .
\end{equation}
By introducing two scalar fields $\lambda$ and $\phi$, we may rewrite the action
(\ref{nlFR1}) in a local form,
\begin{equation}
\label{nlFR2}
S= \int d^4 x \sqrt{-g} \left\{ \frac{1}{2\kappa^2} \left( R + \phi F(R)
+ \lambda \left( \Box^k \phi - R\right) \right)
+ \mathcal{L}_\mathrm{matter} \left( g_{\mu\nu}, \Phi_i \right) \right\} \, .
\end{equation}
Accordingly, the action can be further cast in the following form,
\begin{align}
\label{nlFR2B}
S=& \int d^4 x \sqrt{-g} \left\{ \frac{1}{2\kappa^2} \left( R + \phi_1 F(R)
+ \lambda_k \left( \Box \phi_k - R\right)
+ \sum_{i=1}^{k-1} \lambda_i \left( \Box \phi_i - \phi_{i+1} \right) \right)
+ \mathcal{L}_\mathrm{matter} \left( g_{\mu\nu}, \Phi_i \right) \right\} \nn
=& \int d^4 x \sqrt{-g} \left\{ \frac{1}{2\kappa^2} \left( R + \phi_1 F(R)
 - \sum_{i=1}^k \partial_\mu \lambda_i \partial^\mu \phi_i - \lambda_k R
 - \sum_{i=1}^{k-1} \lambda_i \phi_{i+1} \right)
+ \mathcal{L}_\mathrm{matter} \left( g_{\mu\nu}, \Phi_i \right)
\right\} \, ,
\end{align}
where $\lambda_k=\lambda$ and $\phi_1=\phi$. Furthermore, by
introducing additionally two scalar fields $A$ and $B$, we may
further rewrite the action as follows,
\begin{equation}
\label{nlFR3}
S= \int d^4 x \sqrt{-g} \left\{ \frac{1}{2\kappa^2} \left( A + \phi_1 F(A)
+ B \left( R - A \right)
 - \sum_{i=1}^k \partial_\mu \lambda_i \partial^\mu \phi_i - \lambda_k A
 - \sum_{i=1}^{k-1} \lambda_i \phi_{i+1} \right)
+ \mathcal{L}_\mathrm{matter} \left( g_{\mu\nu}, \Phi_i \right) \right\} \, .
\end{equation}
Upon varying the above action with respect to the scalar field
$A$, we obtain,
\begin{equation}
\label{nlFR4}
B=1 + \phi_1 F'(A) - \lambda_k \, .
\end{equation}
Then we may rewrite the action (\ref{nlFR3}) as follows,
\begin{align}
\label{nlFR5}
S=& \int d^4 x \sqrt{-g} \left\{ \frac{1}{2\kappa^2} \left( \phi_1 F(A)
+ B R - \partial_\mu \left( \phi_1 F'(A) - B\right) \partial^\mu \phi_k
 - \phi_1 F'(A) A
 - \sum_{i=1}^{k-1} \left( \partial_\mu \lambda_i \partial^\mu \phi_i
+ \lambda_i \phi_{i+1} \right) \right) \right. \nn
& \left. + \mathcal{L}_\mathrm{matter} \left( g_{\mu\nu}, \Phi_i \right) \right\} \, .
\end{align}
By performing the following scale transformation of the metric,
\begin{equation}
\label{nlFRa4b}
g_{\mu\nu} = \e^{-\sigma} {\tilde g}_{\mu\nu} \, , \quad
\e^\sigma \equiv B\, ,
\end{equation}
the action~(\ref{nlFRa3}) can be obtained in the Einstein frame,
and it is equal to,
\begin{align}
\label{nlFRa5b}
S_\mathrm{E} =& \int d^4 x \sqrt{-\tilde g} \left[ \frac{1}{2\kappa^2} \left\{
\tilde R - \frac{3}{2} \partial_\mu \sigma \partial^\mu \sigma
+ \e^{-2\sigma} \phi_1 \left( F(A)- F'(A) A\right)
 - \e^{-\sigma} \partial_\mu \left( \phi_1 F'(A) - \e^\sigma \right) \partial^\mu \phi_k
\right. \right. \nn
& \left. \left.
 - \e^{-\sigma} \sum_{i=1}^{k-1} \left( \partial_\mu \lambda_i \partial^\mu \phi_i
+ \lambda_i \phi_{i+1} \right) \right\} + \e^{-2\sigma} \mathcal{L}_\mathrm{matter}
\left( \e^{-\sigma} {\tilde g}_{\mu\nu}, \Phi_i \right) \right] \, .
\end{align}
The above expression indicates that ghost modes occur. First for
the simplicity, we consider the case $k=1$. Then the action
(\ref{nlFRa5b}) is simplified to the following form,
\begin{align}
\label{nlFRa6b}
S_\mathrm{E} =& \int d^4 x \sqrt{-\tilde g} \left[ \frac{1}{2\kappa^2} \left\{
\tilde R - \frac{3}{2} \partial_\mu \sigma \partial^\mu \sigma
 - \e^{-\sigma} F'(A) \partial_\mu \phi_1 \partial^\mu \phi_1
 - \e^{-\sigma} \phi_1 F''(A) \partial_\mu A \partial^\mu \phi_1
+ \partial_\mu \sigma \partial^\mu \phi_1
\right. \right. \nn
& \left. \left.
+ \e^{-2\sigma} \phi_1 \left( F(A)- \phi_1 F'(A) A\right)
\right\}
+ \e^{-2\sigma} \mathcal{L}_\mathrm{matter}
\left( \e^{-\sigma} {\tilde g}_{\mu\nu}, \Phi_i \right) \right]
\, .
\end{align}
The kinetic terms of the scalar fields $\sigma$, $\phi_1$, and $A$
are rewritten as follows,
\begin{equation}
\label{nlFRa7b}
 - \left( \partial_\mu \sigma, \partial_\mu \phi_1, \partial_\mu A \right) K
\left( \begin{array}{c} \partial_\mu \sigma \\ \partial_\mu \phi_1 \\ \partial_\mu A
\end{array} \right) \, , \quad
K \equiv \left( \begin{array}{ccc}
\frac{3}{2} & - \frac{1}{2} & 0 \\
 - \frac{1}{2} & \e^{-\sigma} F'(A) & \frac{1}{2} \e^{-\sigma} \phi_1 F''(A) \\
0 & \frac{1}{2} \e^{-\sigma} \phi_1 F''(A) & 0
\end{array} \right) \, .
\end{equation}
Since the determinant of $K$ is given by,
\begin{equation}
\label{nlFRa8}
\det K = - \frac{3}{8} \e^{-2 \sigma} \phi_1^2 F''(A)^2 \, ,
\end{equation}
in effect $\det K$ is negative and therefore a ghost mode occurs.
Upon varying the action (\ref{nlFRa6b}) with respect to $A$, we
obtain,
\begin{equation}
\label{nlFRa9}
0 = - \e^{-2\sigma} \phi_1 F''(A) A
 - \e^{-\sigma} \phi_1 F''(A) \partial_\mu \sigma \partial^\mu \phi_1
+ \e^{-\sigma} \phi_1 F''(A) \Box \phi_1 \, ,
\end{equation}
which can be solved with respect to $A$ as follows,
\begin{equation}
\label{nlFRa10}
A = \e^{\sigma} \left( - \partial_\mu \sigma \partial^\mu \phi_1
+ \Box \phi_1 \right) \, .
\end{equation}
Then the action (\ref{nlFRa6b}) can be rewritten as follows,
\begin{align}
\label{nlFRa11}
S_\mathrm{E} =& \int d^4 x \sqrt{-\tilde g} \left[
\frac{1}{2\kappa^2} \left\{ \tilde R - \frac{3}{2} \partial_\mu
\sigma \partial^\mu \sigma + \e^{-2\sigma} \phi_1 F \left(
\e^{\sigma} \left( - \partial_\mu \sigma \partial^\mu \phi_1
+ \Box \phi_1 \right) \right) + \partial_\mu \sigma \partial^\mu
\phi_1 \right\} \right. \nn & \left. + \e^{-2\sigma}
\mathcal{L}_\mathrm{matter} \left( \e^{-\sigma}
{\tilde g}_{\mu\nu}, \Phi_i \right) \right]\, .
\end{align}
Thus, the original non-local $F(R)$ gravity with action given in
Eq.~(\ref{nlFR1}) clearly leads to ghost modes. In the next
section we propose some modified models of non-local $F(R)$
gravity which may alleviate the ghosts.

\section{Non-local Gravity Models without Ghost}

In this section, we construct the models of non-local $F(R)$
gravity without ghost modes. As a first example, we consider the
following action,
\begin{equation}
\label{nlFRa1}
S= \int d^4 x \sqrt{-g} \left\{ \frac{1}{2\kappa^2} \left( R - \frac{1}{2}
F(R) \Box^{-1} F(R) \right)
+ \mathcal{L}_\mathrm{matter} \left( g_{\mu\nu}, \Phi_i \right) \right\} \, ,
\end{equation}
which can be rewritten as follows,
\begin{align}
\label{nlFRa2}
S=& \int d^4 x \sqrt{-g} \left\{ \frac{1}{2\kappa^2} \left( R
 - \frac{1}{2} \partial_\mu \phi \partial^\mu \phi - \phi F(R) \right)
+ \mathcal{L}_\mathrm{matter} \left( g_{\mu\nu}, \Phi_i \right)
\right\} \, .
\end{align}
Further, by introducing additionally two scalar field $A$ and $B$,
we may further rewrite the above action as follows,
\begin{equation}
\label{nlFRa3}
S = \int d^4 x \sqrt{-g} \left\{ \frac{1}{2\kappa^2} \left( A
 - \frac{1}{2} \partial_\mu \phi \partial^\mu \phi - \phi F(A)
+ B \left( R - A \right) \right)
+ \mathcal{L}_\mathrm{matter} \left( g_{\mu\nu}, \Phi_i \right)\right\} \, ,
\end{equation}
Upon variation of the action with respect to $A$, we obtain,
\begin{equation}
\label{nlFRa4}
B=1 - \phi F'(A) \, ,
\end{equation}
which can be solved with respect to $A$ as $A=A\left( \phi, B
\right)$. Then we can further rewrite the action (\ref{nlFRa3}) as
follows,
\begin{equation}
\label{nlFRa3BB}
S = \int d^4 x \sqrt{-g} \left\{ \frac{1}{2\kappa^2} \left(A\left( \phi, B \right)
 - \frac{1}{2} \partial_\mu \phi \partial^\mu \phi - \phi F\left(A\left( \phi, B \right)\right)
+ B \left( R - A\left( \phi, B \right) \right) \right)
+ \mathcal{L}_\mathrm{matter} \left( g_{\mu\nu}, \Phi_i \right) \right\} \, .
\end{equation}
By the scale transformation of the metric,
\begin{equation}
\label{nlFRa4BB}
g_{\mu\nu} = \e^{-\sigma} {\tilde g}_{\mu\nu} \, , \quad
\e^\sigma \equiv B\, ,
\end{equation}
the action~(\ref{nlFRa3BB}) can be rewritten in the Einstein frame,
and it is equal to,
\begin{align}
\label{nlFRa5}
S_\mathrm{E} =& \int d^4 x \sqrt{-\tilde g} \left[ \frac{1}{2\kappa^2} \left\{
\tilde R - \frac{3}{2} \partial_\mu \sigma \partial^\mu \sigma
 - \frac{1}{2} \e^{-\sigma} \partial_\mu \phi \partial^\mu \phi
 - U \left( \phi, \sigma \right) \right\}
+ \e^{-2\sigma} \mathcal{L}_\mathrm{matter}
\left( \e^{-\sigma} {\tilde g}_{\mu\nu}, \Phi_i \right) \right] \, , \nn
U \left( \phi, \sigma \right) \equiv &
\left( - \e^{-2\sigma} + \e^{-\sigma} \right) A\left( \phi , \sigma\right)
 - \phi \e^{-2\sigma} F\left(A \left( \phi , \sigma\right) \right) \, .
\end{align}
As it is clear from the kinetic terms of the scalar fields $\phi$
and $\sigma$, the model has no ghosts as long as
$\e^\sigma = B=1 - \phi F'(A) > 0$.

As an extension of the model (\ref{nlFRa1}), we
consider the following model,
\begin{equation}
\label{nlFRb1}
S= \int d^4 x \sqrt{-g} \left\{ \frac{1}{2\kappa^2} \left( G(R) - \frac{1}{2}
F(R) \Box^{-1} F(R) \right)
+ \mathcal{L}_\mathrm{matter} \left( g_{\mu\nu}, \Phi_i \right) \right\} \, ,
\end{equation}
Then instead of (\ref{nlFRa5}) in the Einstein frame, we obtain
the following Einstein frame action,
\begin{align}
\label{nlFRa5bb}
S_\mathrm{E} =& \int d^4 x \sqrt{-\tilde g} \left[ \frac{1}{2\kappa^2} \left\{
\tilde R - \frac{3}{2} \partial_\mu \sigma \partial^\mu \sigma
 - \frac{1}{2} \e^{-\sigma} \partial_\mu \phi \partial^\mu \phi
 - U \left( \phi, \sigma \right) \right\}
+ \e^{-2\sigma} \mathcal{L}_\mathrm{matter}
\left( \e^{-\sigma} {\tilde g}_{\mu\nu}, \Phi_i \right) \right] \, , \nn
U \left( \phi, \sigma \right) \equiv &
 - \e^{-2\sigma} G\left(A \left( \phi , \sigma\right) \right)
+ \e^{-\sigma} A\left( \phi , \sigma\right)
 - \phi \e^{-2\sigma} F\left(A \left( \phi , \sigma\right) \right) \, .
\end{align}
Here $A \left( \phi , \sigma\right)$ is again given by solving
(\ref{nlFRa4}), $\e^\sigma = B=1 - \phi F'(A)$.
Then if $\e^\sigma > 0$, as clear from the kinetic terms, again,
the model has no ghosts.

As another extension of the model (\ref{nlFRa1}), we may consider the
following action,
\begin{equation}
\label{nlFRa6}
S= \int d^4 x \sqrt{-g} \left\{ \frac{1}{2\kappa^2} \left( R - \frac{1}{2}
F(R) \Box^{-k} F(R) \right)
+ \mathcal{L}_\mathrm{matter} \left( g_{\mu\nu}, \Phi_i \right) \right\} \, ,
\end{equation}
which can be rewritten as follows,
\begin{align}
\label{nlFRa7}
S=& \int d^4 x \sqrt{-g} \left\{ \frac{1}{2\kappa^2} \left( R
+ \frac{1}{2} \phi \Box^k \phi - \phi F(R) \right)
+ \mathcal{L}_\mathrm{matter} \left( g_{\mu\nu}, \Phi_i \right) \right\} \nn
= & \int d^4 x \sqrt{-g} \left\{ \frac{1}{2\kappa^2} \left( R
+ \frac{1}{2} \phi_1 \Box \phi_k
+ \sum_{i=1}^{k-1} \lambda_i \left( \Box \phi_i - \phi_{i+1} \right)- \phi F(R) \right)
+ \mathcal{L}_\mathrm{matter} \left( g_{\mu\nu}, \Phi_i \right) \right\} \, ,
\end{align}
Unless $k=1$, this model has a ghost in general. When $k=2$, the
action (\ref{nlFRa7}) takes the following form $\left(
\lambda=\lambda_1, \, \phi=\phi_1 \right)$,
\begin{equation}
\label{nlFRa8b}
S= \int d^4 x \sqrt{-g} \left\{ \frac{1}{2\kappa^2} \left( R
+ \frac{1}{2} \phi \Box \phi_2
+ \lambda \left( \Box \phi - \phi_2 \right)- \phi F(R) \right)
+ \mathcal{L}_\mathrm{matter} \left( g_{\mu\nu}, \Phi_i \right) \right\} \, ,
\end{equation}
By introducing $\lambda$ which is defined as follows,
\begin{equation}
\label{nlFRa9b} \lambda = \frac{1}{2} \eta + \xi \, , \quad \phi_2
= \eta - 2 \xi \, ,
\end{equation}
we may rewrite the action (\ref{nlFRa8b}) as follows,
\begin{equation}
\label{nlFRa10b}
S= \int d^4 x \sqrt{-g} \left\{ \frac{1}{2\kappa^2} \left( R
+ \phi \Box \eta
 - \frac{1}{2} \eta^2 + 2 \xi^2 - \phi F(R) \right)
+ \mathcal{L}_\mathrm{matter} \left( g_{\mu\nu}, \Phi_i \right) \right\} \, .
\end{equation}
Then we may integrate $\xi$ out of the action, and disregard $\xi$
thereafter, thus the resulting action reads,
\begin{equation}
\label{nlFRa11b}
S= \int d^4 x \sqrt{-g} \left\{ \frac{1}{2\kappa^2} \left( R
+ \phi \Box \eta
 - \frac{1}{2} \eta^2 - \phi F(R) \right)
+ \mathcal{L}_\mathrm{matter} \left( g_{\mu\nu}, \Phi_i \right) \right\} \, .
\end{equation}
The kinetic term $\frac{1}{2} \phi \Box \eta$ indicates that a
ghost mode may still occur. In order to avoid this problem, we may
deform the model by adding kinetic terms for $\eta$ and $\phi$ as
follows,
\begin{equation}
\label{nlFRa12}
\tilde S= \int d^4 x \sqrt{-g} \left\{ \frac{1}{2\kappa^2} \left( R
+ \phi \Box \eta + \frac{\alpha}{2} \eta \Box \eta + \frac{\beta}{2} \phi \Box \phi
 - \frac{1}{2} \eta^2 - \phi F(R) \right)
+ \mathcal{L}_\mathrm{matter} \left( g_{\mu\nu}, \Phi_i \right)
\right\} \, ,
\end{equation}
where $\alpha$ and $\beta$ are constants. By the variation of the
action with respect to $\eta$, we obtain,
\begin{equation}
\label{nlFRa13}
0= \Box \phi - \eta + \alpha \Box \eta \, ,
\end{equation}
which can be solved with respect to $\eta$ as follows,
\begin{equation}
\label{lnFRa14}
\eta = \left( 1 - \alpha \Box \right)^{-1} \Box \phi \, .
\end{equation}
By substituting the above expression of $\eta$ into the action
(\ref{nlFRa12}), we obtain,
\begin{equation}
\label{nlFRa15}
\tilde S= \int d^4 x \sqrt{-g} \left\{ \frac{1}{2\kappa^2} \left( R
+ \frac{1}{2} \phi \left( \beta \Box + \left( 1 - \alpha \Box \right)^{-1} \Box^2 \right) \phi
 - \phi F(R) \right)
+ \mathcal{L}_\mathrm{matter} \left( g_{\mu\nu}, \Phi_i \right) \right\} \, .
\end{equation}
Upon variation of the action (\ref{nlFRa15}) with respect to
$\phi$, we obtain,
\begin{equation}
\label{nlFRa16}
0= \left( \beta \Box + \left( 1 - \alpha \Box \right)^{-1} \Box^2 \right) \phi - F(R) \, ,
\end{equation}
which we can solve with respect to $\phi$,
\begin{equation}
\label{nlFRa17}
\phi = \left( \beta \Box + \left( 1 - \alpha \Box \right)^{-1} \Box^2 \right)^{-1} F(R) \, .
\end{equation}
Then by substituting the above expression into
Eq.~(\ref{nlFRa15}), we obtain the following non-local action,
\begin{align}
\label{nlFRa18}
\tilde S=& \int d^4 x \sqrt{-g} \left\{ \frac{1}{2\kappa^2} \left( R
 - \frac{1}{2} F(R) \left( \beta \Box + \left( 1 - \alpha \Box \right)^{-1} \Box^2 \right)^{-1}
F(R) \right)
+ \mathcal{L}_\mathrm{matter} \left( g_{\mu\nu}, \Phi_i \right) \right\} \nn
=& \int d^4 x \sqrt{-g} \left\{ \frac{1}{2\kappa^2} \left( R
 - \frac{1}{2} F(R) \left( 1 - \alpha \Box \right) \Box^{-1}
\left( \beta + \left( 1 - \alpha \beta \right) \Box \right)^{-1} F(R) \right)
+ \mathcal{L}_\mathrm{matter} \left( g_{\mu\nu}, \Phi_i \right) \right\} \, .
\end{align}
In order to determine if the model (\ref{nlFRa18}) or
(\ref{nlFRa12}) has a ghost mode or not, we rewrite the action
(\ref{nlFRa12}) as in Eq.~(\ref{nlFRa3BB}), that is,
\begin{align}
\label{nlFRa12b}
\tilde S= \int d^4 x \sqrt{-g} & \left\{ \frac{1}{2\kappa^2} \left( A\left(\phi, \sigma\right)
 - \partial_\mu \phi \partial^\mu \eta - \frac{\alpha}{2} \partial_\mu \eta \partial^\mu \eta
 - \frac{\beta}{2} \partial_\mu \phi \partial^\mu \phi \right. \right. \nn
& \left. \left.- \frac{1}{2} \eta^2 - \phi F\left( A\left(\phi,
\sigma\right) \right) + \e^\sigma \left( R - A\left(\phi,
\sigma\right) \right) \right) + \mathcal{L}_\mathrm{matter} \left(
g_{\mu\nu}, \Phi_i \right) \right\} \, ,
\end{align}
where $A\left(\phi, \sigma\right)$ is given by solving the
equation $\e^\sigma = 1 - \phi F'(A)$, again. Then by performing
the scale transformation of the metric given in Eq.~(\ref{nlFRa4BB}),
we obtain the action in the Einstein frame,
\begin{align}
\label{nlFRa13b}
{\tilde S}_\mathrm{E} = \int d^4 x \sqrt{-\tilde g} & \left[ \frac{1}{2\kappa^2} \left\{
\tilde R - \frac{3}{2} \partial_\mu \sigma \partial^\mu \sigma
 - \e^{-\sigma} \left( \partial_\mu \phi \partial^\mu \eta
+ \frac{\alpha}{2} \partial_\mu \eta \partial^\mu \eta
+ \frac{\beta}{2} \partial_\mu \phi \partial^\mu \phi \right)
 - U \left( \phi, \sigma \right) \right\} \right. \nn
& \left. + \e^{-2\sigma} \mathcal{L}_\mathrm{matter}
\left( \e^{-\sigma} {\tilde g}_{\mu\nu}, \Phi_i \right) \right] \, , \nn
U \left( \phi, \sigma \right) \equiv &
\left( - \e^{-2\sigma} + \e^{-\sigma} \right) A\left( \phi , \sigma\right)
 - \phi \e^{-2\sigma} F\left(A \left( \phi , \sigma\right) \right)
 - \frac{1}{2} \eta^2 \e^{-2\sigma} \, .
\end{align}
Then as long as the following constraints hold true,
\begin{equation}
\label{nlFRa14}
\alpha + \beta > 0 \, , \quad \alpha \beta > 1\, , \quad \e^\sigma =1 - \phi F'(A) > 0 \, ,
\end{equation}
no ghost modes occur in the theory.

\section{Cosmological Solutions of Ghost-free non-local $F(R)$ Gravity Models}

In this section we shall investigate whether the ghost free
non-local $F(R)$ gravity models discussed in the previous section,
have solutions some cosmological evolutions of interest. Firstly,
let us consider if the model (\ref{nlFRa1}) has the Minkowski
spacetime or the de Sitter spacetime solution. We start with the
form of the action given in Eq.~(\ref{nlFRa3}). Then in addition
to Eq.~(\ref{nlFRa4}), upon varying the action with respect to
$\phi$, $A$, and $g_{\mu\nu}$, we obtain,
\begin{align}
\label{nlFRa15b}
0=& \Box \phi - F\left(A\right) \, , \\
\label{nlFRa16b}
R=& A \, , \\
\label{nlFRa17b}
0=& \frac{1}{2\kappa^2} \left\{ \left(A - \frac{1}{2} \partial_\mu \phi \partial^\mu \phi
 - \phi F\left(A\right) + B \left( R - A \right) \right) g_{\mu\nu}
 - 2 B R_{\mu\nu} + 2 \nabla_\mu \nabla_\nu B - 2 g_{\mu\nu} \Box B \right\} \nn
& + T_{\mathrm{matter}\, \mu\nu} \, .
\end{align}
By using the constants $R_0$ and $\phi_0$, we now assume,
\begin{equation}
\label{nlFRa18b}
R=R_0 \, , \quad R_{\mu\nu} = \frac{R_0}{4} g_{\mu\nu} \, , \quad
\phi = \phi_0 \, ,
\end{equation}
and we neglect the contribution from the perfect matter fluids
hereafter, so $T_{\mathrm{matter}\, \mu\nu} = 0$. Then by using
Eq.~(\ref{nlFRa4}), also $B =1 - \phi F'(A)$, (\ref{nlFRa15b}),
(\ref{nlFRa16b}), and Eq.~(\ref{nlFRa17b}), we obtain,
\begin{equation}
\label{nlFRa19}
B =1 - \phi_0 F' \left( R_0 \right) \, , 0=F\left( R_0 \right) \, , \quad A=R_0 \, , \quad
0= R_0 - \frac{1}{2} R_0 \left( 1 - \phi_0 F' \left( R_0 \right) \right) \, .
\end{equation}
Then if the following conditions for $F(R)$ at $R=R_0$,
\begin{equation}
\label{nlFRa20}
F\left( R_0 \right) = 0 \, , \quad F' \left( R_0 \right) \neq 0 \, ,
\end{equation}
are satisfied, we find,
\begin{equation}
\label{nlFRa21}
\phi_0 = - \frac{1}{F' \left( R_0 \right)} \, , \quad B=2 \, , \quad A=R_0 \, .
\end{equation}
Then if $R_0$, which satisfies the conditions in (\ref{nlFRa20}),
is positive, we have the solution describing the de Sitter
spacetime. On the other hand, if $R_0$ is negative, we obtain the
solution corresponding to the anti-de Sitter spacetime.
Furthermore if $R_0$ vanishes, the solution describes the flat
Minkowski spacetime. The solution (\ref{nlFRa21}) also indicates that,
\begin{equation}
\label{nlFRa22}
\e^\sigma = B=1 - \phi F'(A) =1 - \phi_0 F' \left( R_0 \right) =2 > 0 \, ,
\end{equation}
so this guarantees that no ghost mode occurs upon perturbating
around the background. The general form of $F(R)$ satisfying the
conditions in Eq.~(\ref{nlFRa20}) is given by,
\begin{equation}
\label{nlFRa22b}
F(R) = \left( R - R_0 \right) f(R)\, ,
\end{equation}
where $f(R)$ is an arbitrary function satisfying
$f\left( R_0 \right) \neq 0$. An interesting example is,
\begin{equation}
\label{nlFRa22c}
F(R) = \left( \prod_{i=1}^n \left( R - R_0^{(i)} \right) \right) f(R)\, ,
\end{equation}
Now $f(R)$ satisfies the conditions $f\left( R_0^{(i)} \right)
\neq 0$ $\left( i= 1,2, \cdots, n \right)$, which has several
(anti-)de Sitter solutions corresponding to $R=R_0^{(i)} $.

In order to investigate the stability, we consider the potential $U$
in (\ref{nlFRa5}),
\begin{equation}
\label{nlFRa23}
U \left( \phi, B \right) \equiv
\left( - B^{-2} + B^{-1} \right) A\left( \phi , B\right)
 - \phi \e^{-2\sigma} F\left(A \left( \phi , B\right) \right) \, .
\end{equation}
Then the extrema of the potential $U \left( \phi, B \right)$ are
found by solving the following equations,
\begin{align}
\label{nlFRa24}
0=& \partial_\phi U \left( \phi, B \right)
= \left( - B^{-2} + B^{-1} \right) \partial_\phi A\left( \phi , B\right)
 - \phi B^{-2} F'\left(A \left( \phi , B\right) \right) \partial_\phi A\left( \phi , B\right)
\, . \nn
0=& \partial_B U \left( \phi, B \right)
= \left( 2 B^{-3} - B^{-2} \right) A\left( \phi , B\right)
+ \left( - B^{-2} + B^{-1} \right) \partial_B A\left( \phi , B\right)
+ 2 \phi B^{-3} F'\left(A \left( \phi , B\right) \right) \partial_B A\left( \phi , B\right)\, .
\end{align}
If we assume (\ref{nlFRa21}), we find,
\begin{equation}
\label{nlFRa24}
0 = \partial_\phi U \left( \phi=\phi_0, B=2 \right)
= \partial_B U \left( \phi=\phi_0, B=2 \right) \, .
\end{equation}
Then the (anti-)de Sitter solution in (\ref{nlFRa21}) corresponds
to the extrema of the potential $U \left( \phi, B \right) $. Since
the following holds true,
\begin{align}
\label{nlFRa25}
& \partial_\phi A = - \frac{1-B}{\phi^2 F''(A)} \, , \quad
\partial_B A = - \frac{1}{\phi F''(A)} \, , \quad
\partial_\phi^2 A = \frac{2\left( 1-B \right)}{\phi^3 F''(A)}
- \frac{\left(1-B\right)^2F'''(A)}{\phi^4 F''(A)^3} \, , \nn
& \partial_B^2 A = - \frac{F'''(A)}{\phi^2 F''(A)^3} \, , \quad
\partial_B \partial_\phi A = \partial_\phi \partial_B A
= \frac{1}{\phi^2 F''(A)}- \frac{\left(1-B\right) F'''(A)}{\phi^3 F''(A)^3} \, ,
\end{align}
for the solution in (\ref{nlFRa21}), we find,
\begin{equation}
\label{nlFRa27}
\partial_\phi^2 U = \frac{F'\left( R_0 \right)^3}{2F''\left( R_0 \right)}
 - \frac{ F'\left( R_0 \right)^4 F'''\left( R_0 \right)}{2F''\left( R_0 \right)^3} \, , \quad
\partial_B^2 U = - \frac{R_0}{8} + \frac{F'\left( R_0 \right)}{8F''\left( R_0 \right)} \, ,\quad
\partial_B \partial_\phi U =0 \, .
\end{equation}
Then if the following conditions hold true,
\begin{equation}
\label{nlFRa28}
\frac{F'\left( R_0 \right)}{F''\left( R_0 \right)}
> \frac{ F'\left( R_0 \right)^2 F'''\left( R_0 \right)}{F''\left( R_0 \right)^3} \, , \quad
\frac{F'\left( R_0 \right)}{F''\left( R_0 \right)} > R_0 \, ,
\end{equation}
the solution is stable. For example, we assume
\begin{equation}
\label{nlFRa29} F(R) = c_1 \left( R - R_0 \right) + \frac{c_2}{2}
\left( R - R_0 \right)^2 \, ,
\end{equation}
where $c_1$ and $c_2$ are constant with appropriate dimensions.
Then we find,
\begin{equation}
\label{nlFRa30}
F'\left(R_0 \right) = c_1\, , \quad
F''\left(R_0 \right) = c_2\, , \quad
F''''\left(R_0 \right) = 0\, ,
\end{equation}
and therefore the conditions in Eq.~(\ref{nlFRa28}) are quantified
for the model at hand as follows,
\begin{equation}
\label{nlFRa31}
\frac{c_1}{c_2}>0 \, , \quad \frac{c_1}{c_2}>R_0 \, .
\end{equation}
Therefore if $\frac{c_1}{c_2}>R_0>0$, there is a stable de Sitter solution.

\section{Inflationary Phenomenology of Ghost-free non-Local $F(R)$ Gravity Theory}

Having discussed the essential features of the ghost-free
non-local $F(R)$ theories of gravity, in this section we shall
explore the inflationary phenomenology of the models in the Jordan
frame and in the absence of any perfect matter fluids.
Specifically we shall be interested in the model of Eq.~(\ref{nlFRa1})
or equivalently in the model (\ref{nlFRa2}). We
quote here the gravitational action for convenience, in the
absence of matter fluids, which is,
\begin{align}
\label{modelnewnonlocalghostfreemodel}
S=& \int d^4 x
\sqrt{-g}\frac{1}{2} \left\{ \frac{R}{\kappa^2}
 - \partial_\mu \phi \partial^\mu \phi -2 \phi F(R)
\right\} \, .
\end{align}
In the following we introduce the following notation,
\begin{equation}\label{ffunction}
f(R,\phi)=\frac{R}{\kappa^2}
 - \partial_\mu \phi \partial^\mu \phi -2 \phi F(R)\, ,
\end{equation}
and $f_R$ in the following will stand for,
\begin{equation}
\label{frderivative}
f_R=\frac{\partial f(R,\phi)}{\partial R}=\frac{1}{\kappa^2}
 -2 \phi F'(R)\, ,
\end{equation}
where the prime denotes differentiation with respect to the Ricci
scalar. Also, $f_{\phi}$ in the following will denote,
\begin{equation}
\label{fphiderivative}
f_{\phi}=\frac{\partial f}{\partial \phi}=-2 F(R)\, .
\end{equation}
Assuming that the background metric is a flat
Friedmann-Robertson-Walker (FRW) background with line element,
\begin{equation}
\label{metricfrw}
ds^2 = - dt^2 + a(t)^2 \sum_{i=1,2,3} \left(dx^i\right)^2\, ,
\end{equation}
upon varying the gravitational action
(\ref{modelnewnonlocalghostfreemodel}) with respect to the metric
tensor and the scalar field, we obtain the gravitational equation
of motion, which for the FRW metric read,
\begin{align}
\label{firstfriedmanequation1}
3H^2=& \frac{1}{f_R}\left(\frac{1}{2}\dot{\phi}^2
+\frac{Rf_R-f}{2}-3H\dot{f}_R \right)\, , \\
\label{firstfriedmanequation2}
 -3H^2-2\dot{H}=& \frac{1}{f_R}\left(\frac{1}{2}\dot{\phi}^2
 -\frac{Rf_R-f}{2}+\ddot{f}_R+2H\dot{f}_R \right)\, , \\
\label{firstfriedmanequation3}
0 = & \ddot{\phi}+3H\dot{\phi}-\frac{1}{2}f_{\phi} \, ,
\end{align}
where the ``dot'' in the above equations denotes differentiation
with respect to the cosmic time $t$. For the model
(\ref{modelnewnonlocalghostfreemodel}) the gravitational wave
speed is $c_T=1$ and also the speed of sound is also $c_A=1$
\cite{Hwang:2005hb}. Let us work on the equations of motion, and
by combining Eqs.~(\ref{firstfriedmanequation2}) and
(\ref{firstfriedmanequation3}) we obtain the following,
\begin{equation}
\label{equationofmotionnew}
 -2\dot{H}f_R=\dot{\phi}^2-H\dot{f}_R+\ddot{f}_R\, .
\end{equation}
The inflationary dynamics for models of the form $f(R,\phi)$ have
been worked out in Ref.~\cite{Hwang:2005hb}. The slow-roll indices
are equal to,
\begin{equation}
\label{slowrollparameters}
\epsilon_1=\frac{\dot{H}}{H^2}\, , \quad
\epsilon_2=\frac{\ddot{\phi}}{H\dot{\phi}}\, ,\quad
\epsilon_3=\frac{\dot{f_R}}{2Hf_R}\, ,\quad
\epsilon_4=\frac{\dot{E}}{2HE}\, ,
\end{equation}
where the function $E$ appearing above is defined to be,
\begin{equation}
\label{functionepsilon}
E=\frac{f_R}{\dot{\phi}^2}\left(
\dot{\phi^2}+\frac{3\dot{f}_R}{2f_R}\right)\, .
\end{equation}
The scalar perturbations for the $f(R,\phi)$ theory at hand result
to the following spectral index, which is,
\begin{equation}
\label{spectralindex}
n_s=1+\frac{2 \left( 2\epsilon_1-\epsilon_2+\epsilon_3-\epsilon_4\right)}{1+\epsilon_1}\, ,
\end{equation}
and the tensor-to-scalar ratio is in the case at hand,
\begin{equation}
\label{tensortoscalarration}
r=16 |\epsilon_1-\epsilon_3|\, .
\end{equation}
It is easy to understand that the inflationary phenomenology of
the model (\ref{modelnewnonlocalghostfreemodel}) will highly
depend on the $F(R)$ gravity model. In the following sections we
shall investigate two models of interest, the power law $F(R)$
gravity, in which case $F(R)\sim R^n$ for general $n$, and a
specific case of power-law $F(R)$ gravity, the $R^2$ gravity. In
both cases we shall assume that the slow-roll condition for the
Hubble rate holds true, during the inflationary era, that is,
\begin{equation}
\label{slowrollhubblerate}
\dot{H}\ll H^2\, ,
\end{equation}
while the scalar field will be assumed to satisfy the general
slow-roll evolution,
\begin{equation}
\label{constnatroll}
\ddot{\phi}\ll H\dot{\phi}\, .
\end{equation}
However, for convenience and in order to cover simultaneously all
the possible dynamical evolutions for the scalar field, we shall
assume that the scalar field evolves as,
\begin{equation}
\label{constantroll}
\ddot{\phi}=3\beta H\dot{\phi}\, ,
\end{equation}
so when $\beta \neq 0$, the scalar field obeys the constant-roll
evolution rules \cite{Martin:2012pe,Nojiri:2017qvx}, and when
$\beta=0$ the scalar field evolves in the standard slow-roll way.
Note that the slow-roll conditions for the Hubble rate
(\ref{slowrollhubblerate}) do not affect directly, or are not
related to the evolution of the scalar field. However, the two
evolutions are related once the equations of motion are used, at a
fundamental level though, the evolution conditions are unrelated.

\subsection{General non-local Ghost Free Power-law $F(R)$ Gravity Inflationary Phenomenology}

Let us firstly consider the case that the $F(R)$ gravity appearing
in the action of Eq.~(\ref{modelnewnonlocalghostfreemodel}) has
the general power-law form,
\begin{equation}
\label{powerlawfr}
F(R)=-\alpha R^n\, ,
\end{equation}
with $n$ being any number $1<n<2$ except for $n \neq 2$, which
will be discussed in the next subsection, and $\alpha$ being a
dimensionful parameter of the theory. In this case, by using the
slow-roll condition for the scalar field (\ref{constantroll}), the
equation of motion (\ref{firstfriedmanequation3}) for the scalar
field yields,
\begin{equation}
\label{eqnmotionforscalarnewpowerlaw}
3H(\beta+1)\dot{\phi}+\alpha R^n=0\, ,
\end{equation}
and due to the fact that the Ricci scalar for the FRW metric is
$R=12H^2+6\dot{H}$, in conjunction with the slow-roll condition
(\ref{slowrollhubblerate}), the Ricci scalar during the
inflationary era is $R\sim 12H^2$, so we may algebraically solve
the equation (\ref{eqnmotionforscalarnewpowerlaw}) with respect to
$\dot{\phi}$ and we obtain,
\begin{equation}
\label{solutonfordotphi}
\dot{\phi}\simeq \frac{\gamma H^{2n-1}}{\beta+1}\, ,
\end{equation}
where $\gamma$ stands for,
\begin{equation}
\label{gammadefinitiolocal}
\gamma=\frac{12^n\alpha}{3}\, .
\end{equation}
The slow-roll conditions for the Hubble rate can help us to obtain
approximate relations for the evolution of $\phi$, expressed in
terms of the Hubble rate. In this way, we may solve the equations
of motion in order to have an approximate expression for the
Hubble rate, and eventually discover the phenomenological
implications of the power-law model at hand. Unless we follow this
research line, the solution of the cosmological equations cannot
be obtained in an analytic way, and the task finding the
inflationary dynamics is impossible.

For the model at hand we have,
\begin{equation}
\label{frfunctionderivativepowerlaw}
f_R=\frac{1}{\kappa^2}+2n\phi \alpha R^{n-1}\, ,
\end{equation}
and due to the fact that during the inflationary era, the second
term overwhelms the evolution (recall $n>1$), we have
approximately,
\begin{equation}
\label{approximaderivativefR}
f_R\sim 2n \phi \alpha R^{n-1}\, .
\end{equation}
Then, the first Friedman equation, namely Eq.~(\ref{firstfriedmanequation1})
for the model at hand reads,
\begin{equation}
\label{firstfriednmannequationpowerlaw}
6H^2n\phi \alpha R^{n-1}\simeq \frac{\alpha^2R^{2n}}{2\cdot 3^2 H^2
(\beta+1)^2}+\phi \alpha (n-1)R^n-3H\left( 2n\dot{\phi}\alpha
R^{n-1}+2n (n-1)\phi \alpha R^{n-2}\dot{R}\right)\, .
\end{equation}
The last term $\sim \dot{R}$ is subdominant, therefore at leading
order it can be omitted. Therefore, the resulting equation can be
rewritten as,
\begin{equation}
\label{resultingequationforphipowerlaw}
\phi \left(6H^2\alpha (-n+2) \right)\simeq
\frac{\alpha^2R^{n+1}}{2\cdot 3^2 H^2 (\beta+1)^2}+\frac{2n\alpha^2
R^n}{\beta+1}\, .
\end{equation}
By using the approximation $R\sim 12H^2$, we finally obtain,
\begin{equation}
\label{dagonia}
\phi \sim -A H^{2n-2}\, ,
\end{equation}
where $A$ is,
\begin{equation}
\label{alphaparameter}
A=\frac{1}{6\alpha (-2+n)}\left(\frac{12^{n+1} \alpha^2}
{2\cdot 3^2(\beta+1)^2 }+\frac{2\left(12\right)^n n\alpha^2}{\beta+1}
\right)\, .
\end{equation}
Having Eqs.~(\ref{dagonia}) and (\ref{solutonfordotphi}) at hand,
we can easily express the slow-roll indices
(\ref{slowrollparameters}) in terms of the Hubble rate, always
having in mind the slow-roll conditions
(\ref{slowrollhubblerate}), so we obtain the following simplified
solutions,
\begin{align}
\label{slowrollparameterspowerlawmodel}
\epsilon_1=& \frac{\dot{H}}{H^2}\, , \quad \epsilon_2=3\beta\, , \quad
\epsilon_3\simeq \frac{2^{2 n-1} 3^{n-1} \alpha}{A (\beta
+1)}+\frac{n \dot{H}}{H^2}-\frac{\dot{H}}{H^2}\, , \nonumber \\
\epsilon_4\simeq& -\frac{12 \alpha  n^2 \dot{H}}{H^2 (A-6 \alpha n)}
+\frac{2 A n \dot{H}}{H^2 (A-6 \alpha  n)}+\frac{12 \alpha  n
\dot{H}}{H^2 (A-6 \alpha n)}-\frac{2 A \dot{H}}{H^2 (A-6 \alpha n)} \nonumber \\
& -\frac{2^{4-2 n} 3^{2-n} n^2 \dot{H}^2 A }{H^4 (A-6 \alpha n)}
+\frac{2^{3-2 n} 3^{3-n} n^2 \dot{H}^2 A}{H^4 (A-6 \alpha n)}
 -\frac{2^{3-2 n} 3^{3-n} n \dot{H}^2 A}{H^4 (A-6 \alpha  n)} \nonumber \\
& -\frac{2^{4-2 n} 3^{2-n} n^3 \dot{H}^2 A}{H^4 (A-6 \alpha n)}
+\frac{2^{4-2 n} 3^{2-n} n^2 \dot{H}^2 A \beta}{H^4 (A-6 \alpha n)}
 -\frac{2^{3-2 n} 3^{3-n} n \dot{H}^2 A \beta}{H^4 (A-6 \alpha n)} \nonumber \\
& +\frac{2^{3-2 n} 3^{3-n} n^2 \dot{H}^2 A \beta}{H^4 (A-6 \alpha  n)}
 -\frac{2^{4-2 n} 3^{2-n} n^3 \dot{H}^2 A \beta}{H^4 (A-6 \alpha  n)}\, .
\end{align}
As it is obvious, the slow-roll indices $\epsilon_3$ and
$\epsilon_4$ can be expressed as a function of the slow-roll index
$\epsilon_1$, so these two can be further written as,
\begin{align}
\label{slowrollparameterspowerlawmodellasttwoindices}
\epsilon_3\simeq& \frac{2^{2 n-1} 3^{n-1} \alpha}{A (\beta
+1)}+n \epsilon_1-\epsilon_1\, , \nonumber \\
\epsilon_4\simeq& -\frac{12 \alpha  n^2 \epsilon_1}{A-6 \alpha n}
+\frac{2 A n \epsilon_1}{A-6 \alpha  n}
+\frac{12 \alpha n \epsilon_1}{A-6 \alpha  n}
 -\frac{2 A \epsilon_1}{A-6 \alpha n} \nonumber \\
& -\frac{2^{4-2 n} 3^{2-n} n^3 \epsilon_1^2 A}{A-6 \alpha n}
+\frac{2^{4-2n} 3^{2-n} n^2 \epsilon_1^2 A}{A-6 \alpha n}
+\frac{2^{3-2 n} 3^{3-n} n^2 \epsilon_1^2 A}{A-6 \alpha n}
 -\frac{2^{3-2 n} 3^{3-n} n \epsilon_1^2 A}{A-6 \alpha n} \nonumber \\
& -\frac{2^{3-2 n} 3^{3-n} n \epsilon_1^2 A \beta}{A-6 \alpha n}
 -\frac{2^{4-2 n} 3^{2-n} n^3 \epsilon_1^2 A \beta}{A-6 \alpha n}
+\frac{2^{4-2 n} 3^{2-n} n^2 \epsilon_1^2 A \beta}{A-6 \alpha n}
+\frac{2^{3-2 n} 3^{3-n} n^2 \epsilon_1^2 A \beta}{A-6 \alpha  n} \, .
\end{align}
Eventually, what remains is to find the approximate cosmological
evolution generated by the non-local power-law $F(R)$ gravity
model at hand, in the slow-roll approximation. So by substituting
the evolution for $\phi$ and $\dot{\phi}$ from Eqs.~(\ref{dagonia}) and
(\ref{solutonfordotphi}) in the differential
equation (\ref{equationofmotionnew}), we may obtain a differential
equation for the Hubble rate, and by solving it, we can find the
cosmological evolution during the inflationary era for the model
at hand. So at leading order, in view of Eqs.~(\ref{dagonia}) and
(\ref{solutonfordotphi}), the differential equation
(\ref{equationofmotionnew}) at leading order reads,
\begin{equation}
\label{diffeqnfriedmanneqn1}
 -4\dot{H}\phi n \alpha R^{n-1}\simeq
 \frac{\gamma^2R^{2n-1}}{12^{n-1}(\beta+1)^2}-H2n\dot{\phi}\alpha
 R^{n-1}+2n\dot{\phi} \alpha (n-1)R^{n-2}\dot{R}\, ,
\end{equation}
so by using the approximation $R\sim 12H^2$ which holds true
during the inflationary era, and by keeping the dominant terms, we
finally obtain the following approximate analytic solution,
\begin{equation}
\label{hubblesolution} H(t)=\frac{\alpha  (\beta +1) 2^{2 n} 3^n n
(A (-\beta )-A-\gamma +\gamma  n)}{3 \gamma ^2 t}\, ,
\end{equation}
where $A$ and $\gamma$ are defined in Eqs.~(\ref{alphaparameter})
and (\ref{gammadefinitiolocal}). Note that the above evolution
does not necessarily describe an inflationary evolution, but as we
shall demonstrate, for the values of the parameters that guarantee
the phenomenological viability of the model, the evolution is as
expected an inflationary evolution.

Let us proceed to investigate the phenomenological viability of
the model. The slow-roll indices are easily obtained since these
are expressed in terms of the slow-roll index $\epsilon_1$, which
for the Hubble rate (\ref{hubblesolution}) reads,
\begin{equation}
\label{hubbleslowrollone}
\epsilon_1=-\frac{\gamma ^2 2^{-2 n} 3^{1-n}}{\alpha  (\beta +1) n
(A (-\beta )-A+\gamma -\gamma  n)}\, ,
\end{equation}
and accordingly, the observational indices $n_s$ and $r$ appearing
in Eqs.~(\ref{spectralindex}) and (\ref{tensortoscalarration}) can
easily be calculated and expressed in terms of the free parameters
of the theory. We shall not quote here the resulting expression
for the spectral index, since it is too lengthy, but the
tensor-scalar-ratio has a particular simple form which is,
\begin{equation}
\label{tensortoscalarpowerlaw}
r=-\frac{24 (n-2) \left(3 (\beta +1)^2 n^3-6 \beta  (\beta +1)
n^2+\left(6 \beta ^2+\beta -3\right) n-4\right)}{n (3 (\beta +1)
n+1) \left(6 \beta +3 (\beta +1) n^2-6 (\beta +1) n+7\right)}\, .
\end{equation}
Also it is notable that the resulting expression of the spectral
index is $\alpha$-independent, and the same applies for the
tensor-to-scalar ratio (\ref{tensortoscalarpowerlaw}). At this
point we can check the viability of the model easily by using
specific values for the free parameters. For simplicity we shall
use the reduced Planck physical units system in which
$\kappa^2=1$.

\begin{figure}[h!]
\centering
\includegraphics[width=18pc]{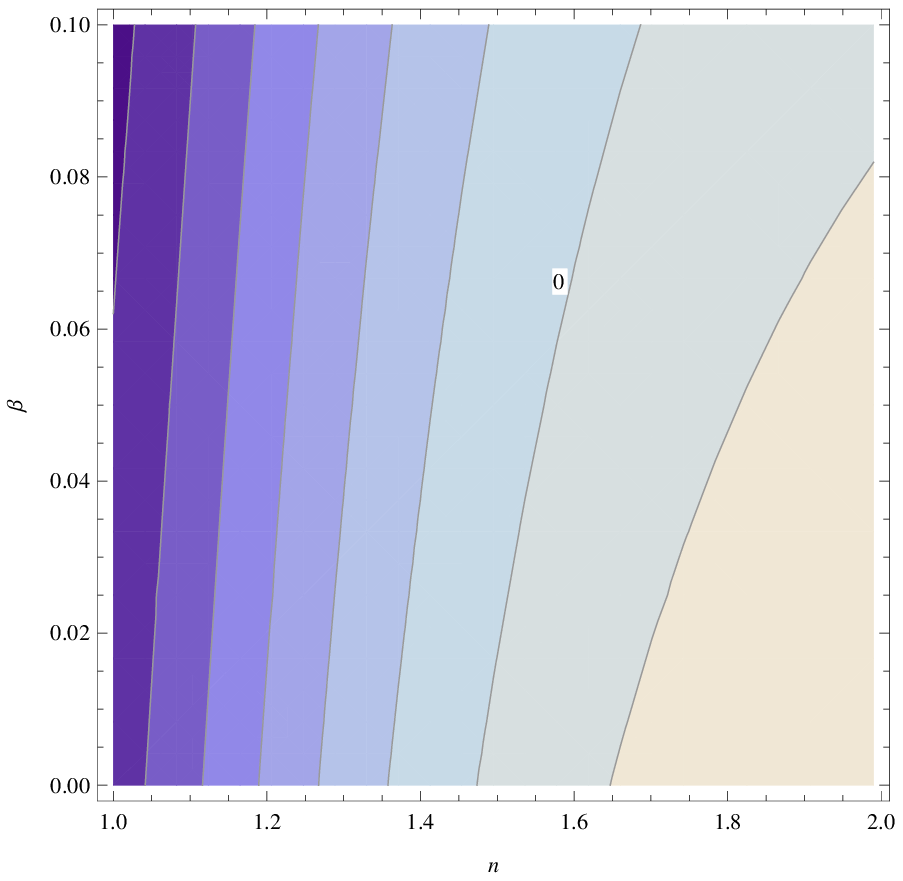}
\includegraphics[width=18pc]{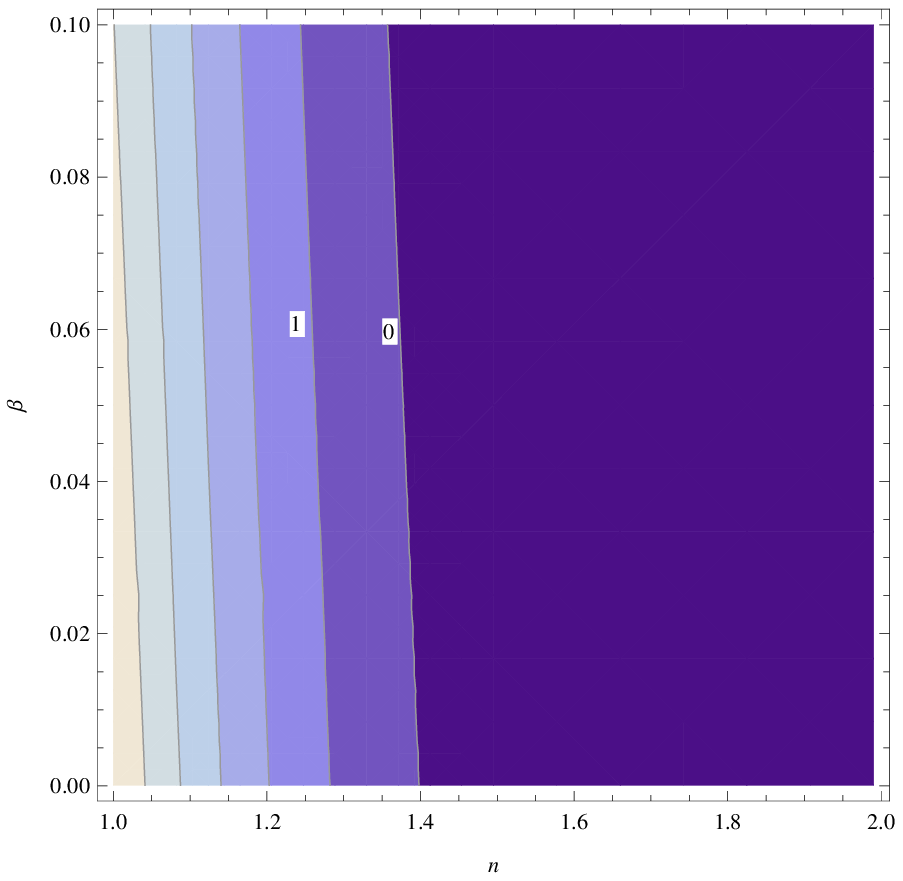}
\caption{Contour plots of the spectral index $n_s$ (left) and of
the tensor-to-scalar ratio $r$ (right) for $n=[0.0001, 1.99]$ and
$\beta=[0.0001, 0.1]$.} \label{plot1}
\end{figure}

A thorough investigation of the free parameters space shows that
the viability of the model comes relatively easy in this power-law
model, however, only in the constant-roll case evolution. Indeed,
by choosing for example $\beta =0.0045$ and $n=1.99$ in reduced
Planck units, we have,
\begin{equation}
\label{observationalresultspowerlaw}
n_s=0.964331\, ,\quad r=0.0341092\, ,
\end{equation}
which are compatible with the latest (2018) Planck constraints
\cite{Akrami:2018odb} on the spectral index and the
tensor-to-scalar ratio, which are,
\begin{equation}
\label{observationaldatanewresults}
n_s=0.962514\pm 0.00406408\, ,\quad r<0.064\, .
\end{equation}
It can be shown that the simultaneous compatibility of the
spectral index and the tensor-to-scalar ratio with the Planck data
occurs for a wide range of values of the free parameters $\beta $
and $n$. This can be seen in Fig.~\ref{plot1} where we present the
plots of the spectral index (left) and of the tensor-to-scalar
ratio (right) in the ranges $n=[0.0001, 1.99]$ and $\beta=[0.0001,0.1]$.
It would be expected that the viability in the constant-roll case
comes for a wider range of values of $\beta$, but on the contrary,
only small values of $\beta$ ($0<\beta <1$) guarantee the
viability of the model. Finally, let us show explicitly that the
evolution given in Eq.~(\ref{hubblesolution}) indeed describes an
inflationary evolution for the allowed values of the parameters.
The Hubble rate (\ref{hubblesolution}) corresponds to the scale
factor,
\begin{align}
\label{scalefactora}
a(t)=&\mathrm{const} \nn
& -\frac{n\left(6 \beta +3 (\beta +1) n^2-6 (\beta +1) n+7\right)
\left(-\frac{n \left(6 \beta +3 (\beta +1) n^2-6 (\beta +1) n+7\right)}{3(n-2)} -1 \right)}
{3 (n-2)}
t^{-\frac{n \left(6 \beta +3 (\beta +1) n^2-6 (\beta +1) n+7\right)}
{3 (n-2)} - 2} \, ,
\end{align}
and by choosing  $\beta =0.0045$ and $n=1.99$ in reduced Planck
units, for which values the viability of the model is ensured, we
have,
\begin{equation}\label{secondderivaitvescalefactor}
\ddot{a}\sim 213117\times t^{460.146}\, ,
\end{equation}
so $\ddot{a}>0$ and we definitely have an inflationary solution.
This result occurs for a wide range of parameter values and an
inflationary solution is guaranteed for values of $\beta$ and $n$
for which $\ddot{a}>0$.


\subsection{Non-local Ghost-free $R^2$ Gravity Inflationary Phenomenology}

Having discussed the non-local power-law $F(R)$ gravity model, in
this section we shall investigate the $R^2$ gravity case. We
discriminate it from the power-law $F(R)$ gravity case, because in
the $R^2$ case, the gravitational equations are more easy to
study, and thus more accurate results may be obtained. This
discrimination between the two cases can be found in the context
of pure $F(R)$ gravity, see the discussion in Ref.~\cite{Odintsov:2019evb}.
In the $R^2$ case, the $F(R)$ gravity
function appearing in the action of Eq.~(\ref{modelnewnonlocalghostfreemodel})
has the form,
\begin{equation}
\label{powerlawfrsq}
F(R)=-\alpha R^2\, ,
\end{equation}
assuming that the constant-roll conditions hold true for the
scalar field (\ref{constantroll}) and the slow-roll condition
holds true for the Hubble rate (\ref{slowrollhubblerate}), we have
for the scalar field equation of motion,
\begin{equation}
\label{eqnmotionforscalarnewpowerlawsq}
3H(\beta+1)\dot{\phi}+\alpha R^2=0\, ,
\end{equation}
and since the Ricci scalar during the
inflationary era is $R\sim 12H^2$, so solve
the equation (\ref{eqnmotionforscalarnewpowerlawsq}) with respect to
$\dot{\phi}$ and we obtain,
\begin{equation}
\label{solutonfordotphi1}
\dot{\phi}\simeq -\frac{\gamma H^{3}}{\beta+1}\, ,
\end{equation}
where $\gamma$ stands for,
\begin{equation}
\label{gammadefinitiolocal}
\gamma=\frac{12^2\alpha}{3}\, .
\end{equation}
Following the method of the previous section, we substitute
$\dot{\phi}$ from Eq.~(\ref{solutonfordotphi1}) into equation
(\ref{firstfriedmanequation1}), and using the fact that during
inflation,
\begin{equation}
\label{approximaderivativefRsq}
f_R\sim 4 \phi \alpha R\, ,
\end{equation}
we easily obtain $\phi$ as a function of the Hubble rate, which is,
\begin{equation}
\label{resultingequationforphipowerlawsq}
\phi \sim A H^2 \, ,
\end{equation}
where $A$ in this case is,
\begin{equation}
\label{alphaparametersq}
A=\frac{\frac{12^4 \alpha ^2}{2\cdot 9 (\beta +1)^2}+\frac{12^4 \alpha ^2}{3}}
{12^2 \alpha +12 \alpha }\, .
\end{equation}
Having Eqs.~(\ref{resultingequationforphipowerlawsq}) and
(\ref{solutonfordotphi1}) at hand, we can easily express the
slow-roll indices (\ref{slowrollparameters}) in terms of the
Hubble rate, as follows, solutions,
\begin{align}
\label{slowrollparameterspowerlawmodelsq}
& \epsilon_1=\frac{\dot{H}}{H^2}\, , \quad \epsilon_2=3\beta\, ,\quad
\epsilon_3\simeq \frac{\Dot{H}}{H^2}-\frac{\gamma }{2 A
(\beta +1)} \, , \nonumber \\
& \epsilon_4\simeq -\frac{288 \alpha  A
\beta  \Dot{H}^2}{\gamma  (72 \alpha +A) H^4}-\frac{288 \alpha A
\Dot{H}^2}{\gamma  (72 \alpha +A) H^4}+\frac{2 A \Dot{H}}{(72
\alpha +A) H^2}+\frac{144 \alpha  \Dot{H}}{(72 \alpha +A) H^2} \, .
\end{align}
As it is obvious, the slow-roll indices $\epsilon_3$ and
$\epsilon_4$ can be expressed in this case too as a function of the slow-roll index
$\epsilon_1$, so we rewrite these two as follows,
\begin{equation}
\label{slowrollparameterspowerlawmodellasttwoindicessq}
\epsilon_3\simeq -\frac{\gamma }{2 A (\beta +1)}+\epsilon_1 \, , \quad
\epsilon_4\simeq -\frac{288 \alpha  A \beta
\epsilon_1^2}{\gamma  (72 \alpha +A)}-\frac{288 \alpha  A
\epsilon_1^2}{\gamma  (72 \alpha +A)}+\frac{2 A \epsilon_1}{72
\alpha +A}+\frac{144 \alpha \epsilon_1}{72 \alpha +A} \, .
\end{equation}
As in the previous section, what now remains is to find the
approximate cosmological evolution realized by the non-local $R^2$
$F(R)$ gravity, so by substituting the evolution for $\phi$ and
$\dot{\phi}$ from Eqs.~(\ref{resultingequationforphipowerlawsq})
and (\ref{solutonfordotphi1}) in the differential equation
(\ref{equationofmotionnew}), we obtain the following differential
equation at leading order,
\begin{equation}
\label{finalsemiapproximation1sq}
B H(t)^2-8\ 24 \alpha  \gamma  H(t)^4 \Dot{H}+4\ 12 \alpha  \gamma
H(t)^2\simeq 0\, ,
\end{equation}
which is obviously different from the one corresponding to the
power-law case appearing in Eq.~(\ref{finalsemiapproximation1}).
Note that $B$ in Eq.~(\ref{finalsemiapproximation1sq}) is equal to,
\begin{equation}
\label{bdefintion}
B=\frac{12^4 \alpha ^2}{9}\, .
\end{equation}
The differential equation (\ref{finalsemiapproximation1sq}) has the following
analytic solution,
\begin{equation}
\label{hubblesolutionsq}
H(t)=\frac{\sqrt[3]{192 \alpha  \gamma  \Lambda +B t
+48 \alpha  \gamma  t}}{4 \sqrt[3]{\alpha } \sqrt[3]{\gamma }}\, ,
\end{equation}
where $\Lambda$ is an integration constant. Obviously the solution
(\ref{hubblesolutionsq}) is different from the one corresponding
to the power-law case of the previous section, namely Eq.~(\ref{hubblesolution}).
The solution (\ref{hubblesolutionsq}) is
particularly interesting, since it describes a non-singular
cosmology, for $\Lambda>0$, or a cosmology which leads to a future
Type III future singularity if $\Lambda<0$, see Ref.~\cite{Nojiri:2005sx}
for the classification of future
singularities. In this paper we shall consider the phenomenology
of the cosmological evolution (\ref{hubblesolutionsq}) in the
context of non-local $R^2$ gravity, but it is surely worth
investigating from which pure $F(R)$ gravity this cosmological
evolution can be generated, and if it is viable, but we defer this
task to a future work. Focusing on the non-local $R^2$ model at
hand, by combining Eqs.~(\ref{slowrollparameterspowerlawmodellasttwoindicessq})
and (\ref{hubblesolutionsq}), also by solving the equation
$\epsilon_1(t_f)=1$ and by expressing the horizon crossing time,
as a function of the $e$-foldings number and the final time $t_f$,
we obtain the analytic form of the slow-roll indices in terms of
the free parameters of the model,
\begin{align}
\label{slowrollparameterspowerlawmodellasttwoindicessqnew}
\epsilon_1=& \frac{\alpha ^2}{4 \alpha ^2 N-1}\, , \quad
\epsilon_2=3\beta\, , \nonumber \\
\epsilon_3\simeq& \alpha ^2
\left(\frac{1}{4 \alpha ^2 N-1}-\frac{13 (\beta +1)}{4 \alpha^2
\left(6 \beta ^2+12 \beta +7\right)}\right) \, , \nonumber \\
\epsilon_4\simeq& \frac{2 \alpha ^2 A}{(72 \alpha +A) \left(4 \alpha ^2 N-1\right)}
+\frac{144 \alpha ^3}{(72 \alpha +A) \left(4 \alpha ^2
N-1\right)} \nonumber \\
& -\frac{288 \alpha ^5 A \beta }{\gamma  (72 \alpha +A) \left(4 \alpha ^2 N-1\right)^2}
 - \frac{288 \alpha ^5 A}{\gamma  (72 \alpha +A) \left(4 \alpha ^2 N-1\right)^2} \, ,
\end{align}
which are evaluated at the time instance of the first horizon
crossing. Accordingly the spectral index of the primordial scalar
curvature perturbations reads,
\begin{equation}
\label{spectralindexofprimordialscalarcurvpert}
n_s=\frac{2 \left(-\frac{\gamma }{2 A \beta +2 A}+\frac{72 \alpha
^3}{(72 \alpha +A) \left(4 \alpha ^2 N-1\right)}+\frac{A \left(288
\alpha ^5 (\beta +1)-\alpha ^2 \gamma +4 \alpha ^4 \gamma
N\right)}{\gamma  (72 \alpha +A) \left(1-4 \alpha ^2 N\right)^2}-3
\beta \right)}{\frac{\alpha ^2}{4 \alpha ^2 N-1}+1}+1\, ,
\end{equation}
while the tensor-to-scalar ratio reads,
\begin{equation}
\label{tensortoscalarrsquaremodel}
r=\frac{52 (\beta +1)}{6 \beta ^2+12 \beta +7}\, ,
\end{equation}
so it is $\beta$ dependent. A direct substitution $\beta=0$ in the
tensor-to-scalar ratio results to $r=7.42857$ which in turn
indicates that the slow-roll case of the $R^2$ non-local $F(R)$
gravity model yields phenomenologically non-viable results. Thus
we concentrate on the constant-roll case with $\beta\neq 0$. In
fact, the 2018 Planck constraint on the tensor-to-scalar ratio,
namely $r<0.064$ can be satisfied when $\beta>140$, so the
viability for this model comes for abnormally large values of
$\beta$. For example, if we work in reduced Planck units with
$\kappa^2=1$, if we choose $N=60$, $\beta=150$ and
$\alpha=0.06466797383519735$, we have,
\begin{equation}
\label{observationalresultspowerlaw}
n_s=0.966\pm 0.00406408,\,\,\,r=0.0573947\, ,
\end{equation}
which are compatible with the latest (2018) Planck constraints
(\ref{observationaldatanewresults}). Although the model is viable
in the constant-roll case, the non-local $R^2$ $F(R)$ gravity
model is deemed rather physically unappealing, due to the
abnormally large values of the parameter $\beta$. This result
indicates that the non-local $F(R)$ gravity models are viable in
general, but the viability is model dependent, like in the
ordinary $F(R)$ gravity models of course.

As we already discussed, the cosmology (\ref{hubblesolutionsq}) is
generated by the non-local $R^2$ model, however, it would be
interesting to investigate which pure $F(R)$ gravity can generate
this cosmological evolution, and whether this cosmological
evolution can be phenomenologically viable. We defer this task to
a future work.

\section{Conclusions}

In this paper we studied ghost-free non-local $F(R)$ gravity
models. Our first aim was to demonstrate how ghost degrees of
freedom can arise in these theories, so after having shown that,
we investigated how it is possible to obtain ghost-free non-local
$F(R)$ gravity theories. By appropriately modifying the
gravitational action of non-local $F(R)$ gravity theories, we
demonstrated that ghost-free theories may be obtained, and we
extracted the conditions which must be fulfilled in order to have
ghost-free theories. Also we investigated whether (anti-)de Sitter
and Minkowski spacetime cosmological solutions may arise in the
theory, and we examined the stability conditions for these
theories. Moreover we studied the inflationary phenomenology of
the Jordan frame ghost-free non-local $F(R)$ gravity. We presented
in detail the essential features of the inflationary dynamics of
the model, quantified in the slow-roll indices and the
corresponding spectral index of the primordial curvature
perturbations and the tensor-to-scalar ratio. Then we focused on
two $F(R)$ gravity models, the power law $F(R)$ gravity model
$\sim R^n$ with $1<n<2$, $n\neq 2$ and the $R^2$ model. The reason
for discriminating among the two power-law models is that in the
case of the $R^2$ model, the gravitational equations of motion are
greatly simplified and more accurate analytic results may be
obtained. In both cases we assumed that the slow-roll condition on
the Hubble rate holds true during the inflationary era, namely
$\dot{H}\ll H^2$ and also that the scalar field obeys the general
constant-roll condition $\ddot{\phi}=3\beta H\dot{\phi}$, which is
reduced to the slow-roll case when $\beta=0$. As we have shown,
the power-law non-local $F(R)$ gravity case produces a viable
inflationary era, compatible with the latest (2018) Planck data,
when the constant-roll condition holds true for the scalar field
even for small $\beta $ values. On the other hand, the $R^2$
non-local $F(R)$ gravity model was deemed not appealing, since it
generates a viable inflationary phenomenology compatible with the
observational data only when $\beta$ takes values larger than
$\beta >140$, which is rather not appealing, if not unphysical.
The results are of course model-dependent and a better choice of
the $F(R)$ gravity can in principle enhance the viability of the
model even in the slow-roll evolution case for the scalar field,
but we chose the simplest examples in order to just demonstrate
the argument.

An interesting feature we found is the form of the Hubble rate
during the inflationary era corresponding to the $R^2$ non-local
$F(R)$ gravity model. It is interesting to investigate from which
pure vacuum $F(R)$ gravity this cosmological evolution is
realized, and whether this evolution is a viable cosmological
evolution. Work is in progress towards this research line.

\section*{Acknowledgments}

This work is supported by MINECO (Spain), FIS2016-76363-P, and by
project 2017 SGR247 (AGAUR, Catalonia) (S.D.O). This work is also
supported by MEXT KAKENHI Grant-in-Aid for Scientific Research on
Innovative Areas ``Cosmic Acceleration'' No. 15H05890 (S.N.) and
the JSPS Grant-in-Aid for Scientific Research (C) No. 18K03615
(S.N.).


\begin{thebibliography}{99}

\bibitem{Perlmutter:1998np}
S.~Perlmutter {\it et al.} [Supernova Cosmology Project
Collaboration],
Astrophys.\ J.\ {\bf 517} (1999) 565 doi:10.1086/307221
[astro-ph/9812133].

\bibitem{Riess:1998cb}
A.~G.~Riess {\it et al.} [Supernova Search Team],
Astron.\ J.\ {\bf 116} (1998) 1009 doi:10.1086/300499
[astro-ph/9805201].

\bibitem{Spergel:2003cb}
D.~N.~Spergel {\it et al.} [WMAP Collaboration],
Astrophys.\ J.\ Suppl.\ {\bf 148} (2003) 175 doi:10.1086/377226
[astro-ph/0302209].

\bibitem{Spergel:2006hy}
D.~N.~Spergel {\it et al.} [WMAP Collaboration],
Astrophys.\ J.\ Suppl.\ {\bf 170} (2007) 377 doi:10.1086/513700
[astro-ph/0603449].

\bibitem{Komatsu:2008hk}
E.~Komatsu {\it et al.} [WMAP Collaboration],
Astrophys.\ J.\ Suppl.\ {\bf 180} (2009) 330
doi:10.1088/0067-0049/180/2/330 [arXiv:0803.0547 [astro-ph]].

\bibitem{Komatsu:2010fb}
E.~Komatsu {\it et al.} [WMAP Collaboration],
Astrophys.\ J.\ Suppl.\ {\bf 192} (2011) 18
doi:10.1088/0067-0049/192/2/18 [arXiv:1001.4538 [astro-ph.CO]].

\bibitem{Tegmark:2003ud}
M.~Tegmark {\it et al.} [SDSS Collaboration],
Phys.\ Rev.\ D {\bf 69} (2004) 103501
doi:10.1103/PhysRevD.69.103501 [astro-ph/0310723].

\bibitem{Seljak:2004xh}
U.~Seljak {\it et al.} [SDSS Collaboration],
Phys.\ Rev.\ D {\bf 71} (2005) 103515
doi:10.1103/PhysRevD.71.103515 [astro-ph/0407372].

\bibitem{Eisenstein:2005su}
D.~J.~Eisenstein {\it et al.} [SDSS Collaboration],
Astrophys.\ J.\ {\bf 633} (2005) 560 doi:10.1086/466512
[astro-ph/0501171].

\bibitem{Jain:2003tba}
B.~Jain and A.~Taylor,
Phys.\ Rev.\ Lett.\ {\bf 91} (2003) 141302
doi:10.1103/PhysRevLett.91.141302 [astro-ph/0306046].


\bibitem{Nojiri:2017ncd}
S.~Nojiri, S.~D.~Odintsov and V.~K.~Oikonomou,
Phys.\ Rept.\  {\bf 692} (2017) 1
doi:10.1016/j.physrep.2017.06.001 [arXiv:1705.11098 [gr-qc]].

\bibitem{Nojiri:2010wj}
S.~Nojiri and S.~D.~Odintsov,
Phys.\ Rept.\ {\bf 505} (2011) 59
doi:10.1016/j.physrep.2011.04.001 [arXiv:1011.0544 [gr-qc]].

\bibitem{Nojiri:2006ri}
S.~Nojiri and S.~D.~Odintsov,
eConf C {\bf 0602061} (2006) 06
 [Int.\ J.\ Geom.\ Meth.\ Mod.\ Phys.\ {\bf 4} (2007) 115]
doi:10.1142/S0219887807001928 [hep-th/0601213].




\bibitem{Capozziello:2011et}
S.~Capozziello and M.~De Laurentis,
Phys.\ Rept.\ {\bf 509} (2011) 167
doi:10.1016/j.physrep.2011.09.003 [arXiv:1108.6266 [gr-qc]].

\bibitem{Capozziello:2010zz}
V.~Faraoni and S.~Capozziello,
Fundam.\ Theor.\ Phys.\ {\bf 170} (2010).
doi:10.1007/978-94-007-0165-6

\bibitem{delaCruzDombriz:2012xy}
A.~de la Cruz-Dombriz and D.~Saez-Gomez,
Entropy {\bf 14} (2012) 1717 doi:10.3390/e14091717
[arXiv:1207.2663 [gr-qc]].

\bibitem{Olmo:2011uz}
G.~J.~Olmo,
Int.\ J.\ Mod.\ Phys.\ D {\bf 20} (2011) 413
doi:10.1142/S0218271811018925 [arXiv:1101.3864 [gr-qc]].

\bibitem{Nojiri:2003ft}
S.~Nojiri and S.~D.~Odintsov,
Phys.\ Rev.\ D {\bf 68} (2003) 123512
doi:10.1103/PhysRevD.68.123512
[hep-th/0307288].

\bibitem{Modesto:2017sdr}
L.~Modesto and L.~Rachwal,
Int.\ J.\ Mod.\ Phys.\ D {\bf 26} (2017) no.11,  1730020.
doi:10.1142/S0218271817300208

\bibitem{Belgacem:2017cqo}
E.~Belgacem, Y.~Dirian, S.~Foffa and M.~Maggiore,
JCAP {\bf 1803} (2018) 002
doi:10.1088/1475-7516/2018/03/002
[arXiv:1712.07066 [hep-th]].

\bibitem{Koshelev:2016xqb}
A.~S.~Koshelev, L.~Modesto, L.~Rachwal and A.~A.~Starobinsky,
JHEP {\bf 1611} (2016) 067
doi:10.1007/JHEP11(2016)067
[arXiv:1604.03127 [hep-th]].


\bibitem{Wetterich:1997bz}
C.~Wetterich,
Gen.\ Rel.\ Grav.\  {\bf 30} (1998) 159
doi:10.1023/A:1018837319976
[gr-qc/9704052].

\bibitem{Deser:2007jk}
S.~Deser and R.~P.~Woodard,
Phys.\ Rev.\ Lett.\  {\bf 99} (2007) 111301
doi:10.1103/PhysRevLett.99.111301
[arXiv:0706.2151 [astro-ph]].


\bibitem{Nojiri:2007uq}
S.~Nojiri and S.~D.~Odintsov,
Phys.\ Lett.\ B {\bf 659} (2008) 821
doi:10.1016/j.physletb.2007.12.001 [arXiv:0708.0924 [hep-th]].

\bibitem{ArkaniHamed:2002fu}
N.~Arkani-Hamed, S.~Dimopoulos, G.~Dvali and G.~Gabadadze,
hep-th/0209227.

\bibitem{Nojiri:2010pw}
S.~Nojiri, S.~D.~Odintsov, M.~Sasaki and Y.~l.~Zhang,
Phys.\ Lett.\ B {\bf 696} (2011) 278
doi:10.1016/j.physletb.2010.12.035 [arXiv:1010.5375 [gr-qc]].

\bibitem{Joukovskaya:2007nq}
L.~Joukovskaya,
Phys.\ Rev.\ D {\bf 76} (2007) 105007
doi:10.1103/PhysRevD.76.105007 [arXiv:0707.1545 [hep-th]].

\bibitem{Calcagni:2007ef}
G.~Calcagni, M.~Montobbio and G.~Nardelli,
Phys.\ Lett.\ B {\bf 662} (2008) 285
doi:10.1016/j.physletb.2008.03.024 [arXiv:0712.2237 [hep-th]].

\bibitem{Jhingan:2008ym}
S.~Jhingan, S.~Nojiri, S.~D.~Odintsov, M.~Sami, I.~Thongkool and
S.~Zerbini,
Phys.\ Lett.\ B {\bf 663} (2008) 424
doi:10.1016/j.physletb.2008.04.054 [arXiv:0803.2613 [hep-th]].

\bibitem{Capozziello:2008gu}
S.~Capozziello, E.~Elizalde, S.~Nojiri and S.~D.~Odintsov,
Phys.\ Lett.\ B {\bf 671} (2009) 193
doi:10.1016/j.physletb.2008.11.060 [arXiv:0809.1535 [hep-th]].

\bibitem{Koshelev:2008ie}
N.~A.~Koshelev,
Grav.\ Cosmol.\ {\bf 15} (2009) 220 doi:10.1134/S0202289309030049
[arXiv:0809.4927 [gr-qc]].

\bibitem{Nesseris:2009jf}
S.~Nesseris and A.~Mazumdar,
Phys.\ Rev.\ D {\bf 79} (2009) 104006
doi:10.1103/PhysRevD.79.104006 [arXiv:0902.1185 [astro-ph.CO]].

\bibitem{Deffayet:2009ca}
C.~Deffayet and R.~P.~Woodard,
JCAP {\bf 0908} (2009) 023 doi:10.1088/1475-7516/2009/08/023
[arXiv:0904.0961 [gr-qc]].

\bibitem{Calcagni:2009dg}
G.~Calcagni and G.~Nardelli,
Int.\ J.\ Mod.\ Phys.\ D {\bf 19} (2010) 329
doi:10.1142/S0218271810016440 [arXiv:0904.4245 [hep-th]].

\bibitem{Cognola:2009jx}
G.~Cognola, E.~Elizalde, S.~Nojiri, S.~D.~Odintsov and S.~Zerbini,
Eur.\ Phys.\ J.\ C {\bf 64} (2009) 483
doi:10.1140/epjc/s10052-009-1154-4 [arXiv:0905.0543 [gr-qc]].

\bibitem{Bronnikov:2009az}
K.~A.~Bronnikov and E.~Elizalde,
Phys.\ Rev.\ D {\bf 81} (2010) 044032
doi:10.1103/PhysRevD.81.044032 [arXiv:0910.3929 [hep-th]].

\bibitem{Calcagni:2010ab}
G.~Calcagni and G.~Nardelli,
Phys.\ Rev.\ D {\bf 82} (2010) 123518
doi:10.1103/PhysRevD.82.123518 [arXiv:1004.5144 [hep-th]].

\bibitem{Vernov:2010ui}
S.~Y.~Vernov,
Phys.\ Part.\ Nucl.\ Lett.\ {\bf 8} (2011) 310
doi:10.1134/S1547477111030228 [arXiv:1005.0372 [astro-ph.CO]].



\bibitem{Barnaby:2010kx}
N.~Barnaby,
Nucl.\ Phys.\ B {\bf 845} (2011) 1
doi:10.1016/j.nuclphysb.2010.11.016 [arXiv:1005.2945 [hep-th]].

\bibitem{Dimitrijevic:2019pct}
  I.~Dimitrijevic, B.~Dragovich, A.~S.~Koshelev, Z.~Rakic and J.~Stankovic,
  doi:10.1016/j.physletb.2019.134848
  arXiv:1906.07560 [gr-qc].


\bibitem{Dialektopoulos:2018iph}
  K.~F.~Dialektopoulos, D.~Borka, S.~Capozziello, V.~Borka Jovanovic and P.~Jovanovic,
  Phys.\ Rev.\ D {\bf 99} (2019) no.4,  044053
  doi:10.1103/PhysRevD.99.044053
  [arXiv:1812.09289 [astro-ph.GA]].



\bibitem{Calmet:2018rkj}
  X.~Calmet, B.~K.~El-Menoufi, B.~Latosh and S.~Mohapatra,
  Eur.\ Phys.\ J.\ C {\bf 78} (2018) no.9,  780
  doi:10.1140/epjc/s10052-018-6265-3
  [arXiv:1809.07606 [hep-th]]

\bibitem{Bahamonde:2017sdo}
  S.~Bahamonde, S.~Capozziello and K.~F.~Dialektopoulos,
  Eur.\ Phys.\ J.\ C {\bf 77} (2017) no.11,  722
  doi:10.1140/epjc/s10052-017-5283-x
  [arXiv:1708.06310 [gr-qc]].

\bibitem{Elizalde:2011su}
E.~Elizalde, E.~O.~Pozdeeva and S.~Y.~Vernov,
Phys.\ Rev.\ D {\bf 85} (2012) 044002
doi:10.1103/PhysRevD.85.044002 [arXiv:1110.5806 [astro-ph.CO]].

\bibitem{Bamba:2012ky}
K.~Bamba, S.~Nojiri, S.~D.~Odintsov and M.~Sasaki,
Gen.\ Rel.\ Grav.\ {\bf 44} (2012) 1321
doi:10.1007/s10714-012-1342-7 [arXiv:1104.2692 [hep-th]].



\bibitem{Deser:2013uya}
S.~Deser and R.~P.~Woodard,
JCAP {\bf 1311} (2013) 036
doi:10.1088/1475-7516/2013/11/036
[arXiv:1307.6639 [astro-ph.CO]].


\bibitem{Deser:2019lmm}
S.~Deser and R.~P.~Woodard,
JCAP {\bf 1906}, 034 (2019)
doi:10.1088/1475-7516/2019/06/034
[arXiv:1902.08075 [gr-qc]].




\bibitem{Maggiore:2014sia}
M.~Maggiore and M.~Mancarella,
Phys.\ Rev.\ D {\bf 90} (2014) no.2,  023005
doi:10.1103/PhysRevD.90.023005
[arXiv:1402.0448 [hep-th]].


\bibitem{Chan:2012jj}
K.~C.~Chan, R.~Scoccimarro and R.~K.~Sheth,
Phys.\ Rev.\ D {\bf 85} (2012) 083509
doi:10.1103/PhysRevD.85.083509
[arXiv:1201.3614 [astro-ph.CO]].





\bibitem{Zhang:2011uv}
Y.~l.~Zhang and M.~Sasaki,
Int.\ J.\ Mod.\ Phys.\ D {\bf 21} (2012) 1250006
doi:10.1142/S021827181250006X [arXiv:1108.2112 [gr-qc]].



\bibitem{Akrami:2018odb}
Y.~Akrami {\it et al.} [Planck Collaboration],
arXiv:1807.06211 [astro-ph.CO].



\bibitem{Hwang:2005hb}
J.~c.~Hwang and H.~Noh,
Phys.\ Rev.\ D {\bf 71} (2005) 063536
doi:10.1103/PhysRevD.71.063536 [gr-qc/0412126].



\bibitem{Martin:2012pe}
  J.~Martin, H.~Motohashi and T.~Suyama,
  Phys.\ Rev.\ D {\bf 87} (2013) no.2,  023514
  doi:10.1103/PhysRevD.87.023514
  [arXiv:1211.0083 [astro-ph.CO]].


\bibitem{Nojiri:2017qvx}
  S.~Nojiri, S.~D.~Odintsov and V.~K.~Oikonomou,
  Class.\ Quant.\ Grav.\  {\bf 34} (2017) no.24,  245012
  doi:10.1088/1361-6382/aa92a4
  [arXiv:1704.05945 [gr-qc]].



\bibitem{Odintsov:2019evb}
S.~D.~Odintsov and V.~K.~Oikonomou,
Phys.\ Rev.\ D {\bf 99} (2019) no.10,  104070
doi:10.1103/PhysRevD.99.104070
[arXiv:1905.03496 [gr-qc]].

\bibitem{Nojiri:2005sx}
S.~Nojiri, S.~D.~Odintsov and S.~Tsujikawa,
Phys.\ Rev.\ D {\bf 71} (2005) 063004
doi:10.1103/PhysRevD.71.063004
[hep-th/0501025].





\end{thebibliography}
\end{document}